\documentclass[mathleft
]{an}
\usepackage{graphicx}
\usepackage{times}
\begin{document}

\Pagespan{1}{}
\Yearpublication{2009}%
\Yearsubmission{2009}%
\Month{1}%
\Volume{1}%
\Issue{1}%

\title{Relationship between group sunspot number and Wolf sunspot number}

\author{K.J. Li\inst{1,2} H.F. Liang\inst{3}\fnmsep\thanks{Corresponding author:
  \email{lhf@ynao.ac.cn or lkj@ynao.ac.cn}\newline}
}
\titlerunning{Group sunspot number and sunspot number}
\authorrunning{K.J. Li and H.F. Liang}
\institute{National Astronomical Observatories/Yunnan Observatory,
      CAS, Kunming 650011,   China
\and
Key Laboratory of Solar Activity, National Astronomical Observatories, CAS, Beijing 100012, China
\and
Department of Physics, Yunnan Normal University, Kunming 650093, China}
\received{}
\accepted{}
\publonline{later}


\abstract{Continuous wavelet transform and cross-wavelet transform have been
used to investigate the phase periodicity and synchrony of the
monthly mean Wolf ($R_{z}$) and group ($R_{g}$) sunspot numbers
during the period of June 1795 to December 1995. The Schwabe cycle
is the only one common  period in Rg and Rz, but it is  not
well-defined in case of cycles  5-7  of Rg and  in case of cycles 5
and 6 of $R_{z}$. In fact, the Schwabe period is slightly different in $R_{g}$
and $R_{z}$ before cycle 12, but from cycle 12 onwards it is almost the
same for the two time series. Asynchrony of the two time series is
more obviously seen in cycles 5 and 6 than in the following cycles,
and usually more obviously seen around the maximum time of a cycle
than during the rest of the cycle. $R_{g}$ is found to fit
$R_{z}$ better in both amplitudes and peak epoch during the minimum
time time of a solar cycle than during the maximum time of the
cycle, which should be caused by their different definition, and
around the maximum
time of a cycle,
$R_{g}$ is usually less than $R_{z}$.
Asynchrony of
$R_{g}$ and $R_{z}$ should somewhat agree with different sunspot
cycle characteristics exhibited by themselves.
}

\maketitle
\section{Introduction}
The single but most important index of solar activity has
been the Zurich or Wolf sunspot number (the international sunspot
number or the sunspot number) (Hathaway, Wilson, $\&$ Reichmann
2002;  Hathaway 2010). For more than 100 years the Wolf sunspot number has served
as the primary time series to define solar activity (Hoyt $\&$
Schatten 1998a, 1998b). It has been proven invaluable in studies
of long-term variations in solar activity, especially as related to
terrestrial climate (e.g., Eddy 1976; Hoyt $\&$ Schatten 1997; Li, Gao $\&$ Su
2005; ).

The sunspot number is defined as $R_{z} = k(10g+ f )$, where $k$ is
the normalization constant for a particular observer, $g$ is the
number of sunspot groups, and $f$ the number of individual sunspots
visible over the solar disk. In spite of the apparent arbitrary
nature of this formula, it has been found to correlate extremely
well with other, more physical measures of solar activity such as
sunspot area, 10.7 cm radio flux, X-ray flare frequency, and
magnetic flux (Hathaway, Wilson, $\&$ Reichmann 2002). Monthly
values for $R_{z}$ are available from 1749 onward, but many of
$R_{z}$ values between 1700 and 1850 are clearly based on inaccurate
or missing data. Thus, these data are less accurate than more recent
observations, and the $R_{z}$ values appear to be too large by
25\,--\,50$\%$ prior to 1882 (Hoyt $\&$ Schatten 1995a, b, c, d;
Wilson 1998; Faria et al. 2004). A new parameter, the group sunspot
number ($R_{g}$) was introduced by Hoyt $\&$ Schatten (1998a, 1998b)
as an alternative to the sunspot number. It uses only the number of
sunspot groups but is normalized to make it to agree with the Zurich
sunspot number. With this normalization the group sunspot number is
given by $R_{g} = {{12.08}\over {N}}\sum_{i=1}^{N}(k_{i} g_{i})$,
where $k_{i}$ is the correction factor for observer $i$, $g_{i}$ is
the number of sunspot groups observed by observer $i$, and $N$ the
number of observers used to form the daily value. The group sunspot
number is more self-consistent and less noisy than the  sunspot
number. Daily, monthly, and annual observations were determined
during the period of the years 1610\,--\,1995. The Wolf and group
sunspot numbers are the longest direct instrumental records of solar
activity (Rigozo et al. 2001;
Kane 2002; Ogurtsov et al. 2002; Echer et al. 2005). Hathaway, Wilson, $\&$ Reichmann
(2002) examined the group sunspot number and compared the sunspot
cycle characteristics it exhibits to those exhibited by the Wolf
sunspot number. They found that the Wolf sunspot number continues
to be valuable in capturing characteristics of the recent cycles
that are not quite as well-reflected in the group number, and the
group sunspot number is valuable for capturing the behavior in the
earliest cycles that help to reveal long-term behavior. Ogurtsov et
al. (2002) used the yearly Wolf sunspot number and the yearly group
sunspot number to study long periods of solar activity through
Morlet wavelet analysis and Fourier analysis. Faria et al. (2004)
compared the spectral features of the yearly group sunspot number
and the yearly Wolf sunspot number with the use of multitaper
analysis and Morlet wavelet analysis. Li et al. (2005)
investigated the Schwabe and Gleissberg periods in the Wolf sunspot
number and the group sunspot number. However, no one has
investigated  phase relationship between them up to now, and the
difference between the two caused by their different definition has
not yet been paid enough attention.

According to the modern point of view,  synchronization is an
universal concept in nonlinear sciences (Pikovsky, Rosenblum $\&$
Kurths 2001; Maraun $\&$ Kurths 2004; Romano et al. 2005).  In the present
study, we will use the cross-wavelet transform method to study phase
relation between the monthly mean group sunspot number and the
monthly mean sunspot number and investigate the
influence of their different definition on the relation between them.

\section{Wavelet analyses of the monthly mean Wolf and group sunspot numbers}
\subsection{Data}
The continuous time series of the monthly mean sunspot number
($R_{z}$) is available from January 1749, and the continuous time
series of the monthly mean group sunspot number ($R_{g}$) is
available from June 1795 to December 1995. Figure 1 shows both
$R_{g}$ and $R_{z}$ from June 1795 to December 1995, which are both
downloaded from the NOAA's web
site\footnote{$ftp://ftp.ngdc.noaa.gov/STP/SOLAR_{-}DATA/\\SUNSPOT_{-}NUMBERS$}.
These two kinds of data are used to investigate the phase
relationship between them. As the figure shows, $R_{g}$
obviously differs from $R_{z}$ around the maximum time of a cycle,
implying that they  should be asynchronous.

\begin{figure}
\hskip 10.mm
\includegraphics[width=70mm,height=80mm]{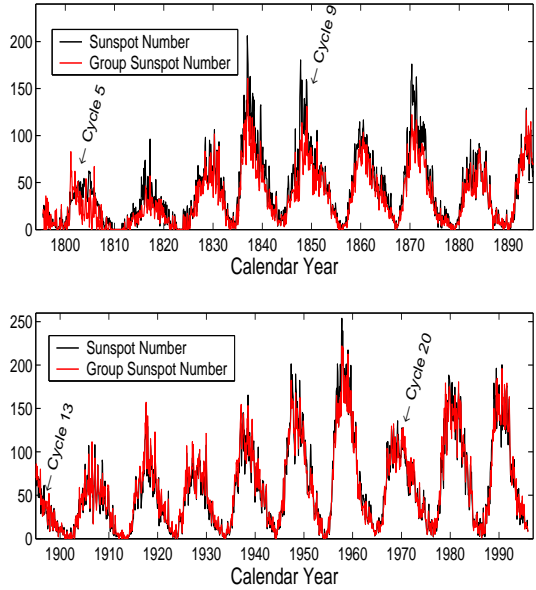}
\caption{The monthly mean group sunspot number (the black line) and
the monthly mean sunspot number (the red line) from June 1795 to
December 1995.
 }
\label{label2}
\end{figure}

\subsection{Continuous wavelet transform}
Wavelet analysis involves a transform from an one-dimensional
time-series to a diffuse two-dimensional time-frequency image for
detecting the localized and (pseudo-) periodic fluctuations by using
the limited time span of the data (Torrence $\&$ Compo 1998; Li et al.
 2005). Here the Morlet continuous wavelet transform (CWT) is
used with its dimensionless frequency $\omega_{0}=6$.
Figure  2 shows the continuous wavelet power
spectra of $R_{g}$ and $R_{z}$. There are evidently common features
in the wavelet power spectra of the two time series. The both have a
large-scale periodicity (about 11 years, namely the Schwabe cycle)
of the highest power, above the 95$\%$ confidence level. However,
the Schwabe cycle does not significantly exist in cycles 5 to 7 for
$R_{g}$  and in cycles 5 and 6 for $R_{z}$.

\begin{figure}
\hskip 10.mm
\includegraphics[width=70mm,height=80mm]{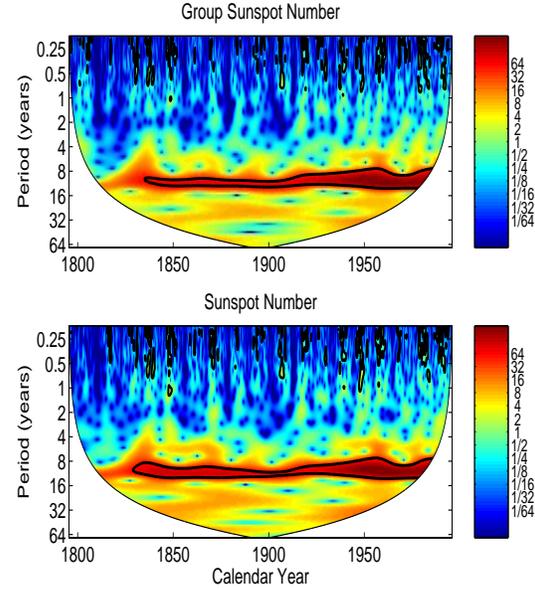}
\caption{The continuous wavelet power spectra of the monthly mean
group sunspot number (top) and the monthly mean sunspot number
(bottom). The thick black contours indicate the 95$\%$ confidence
level, and the white region below the thin solid line indicates the
cone of influence (COI) where edge effects might distort the picture
(for details, see Torrence $\&$ Compo 1998).
 }
\label{label2}
\end{figure}

Figure  3 displays the period length of the Schwabe cycle varying
with time respectively for the two time series. At a certain time
point, the Schwabe period (scale) of  $R_{g}$ (or $R_{z}$) has the
highest spectral power among all considered time scales in its local
wavelet power spectrum, thus the Schwabe period length is determined at a certain time point.
As the figure shows, the Schwabe period for
$R_{g}$ and $R_{z}$ actually differs from each other before cycle
12, but from cycle 12 onwards it is almost the same for the two
time series. If period (frequency) of two time series is different,
the two are asynchronous. The differences of the Schwabe cycle
length for the two time series should lead to a phase asynchrony
between them.

\begin{figure}
\hskip 10.mm
\includegraphics[width=70mm,height=80mm]{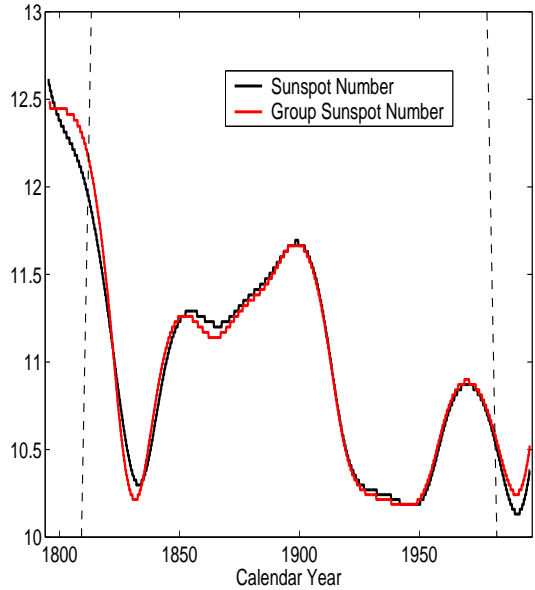}
\caption{The period length of the Schwabe cycle for the monthly mean
group sunspot number (the red line) and that for the monthly mean
sunspot number (the black solid line). The two dashed lines indicate
the COI.
 }
\label{label3}
\end{figure}

\subsection{Cross-wavelet transform}
Cross-wavelet transform (XWT) is an extension of wavelet transforms
to expose the common power and relative phase between two time
series in time-frequency space (Li et al 2009 and references
therein). The cross\,--\,wavelet transform of two time series $X$
and $Y$ is defined as $W^{XY} = W^{X}W^{Y\ast}$, where $\ast$
denotes complex conjugation and $W^{X}$ and $W^{Y}$ are the
continuous wavelet transforms of the individual time series
 (Grinsted et al. 2004). The complex argument $arg(W^{XY})$ can be interpreted as a
local relative phase  between $X$ and $Y$ in time-frequency space,
namely the phase angle difference of $X$ and $Y$. We employ the
codes provided by Grinsted et al.
(2004)\footnote{http://www.pol.ac.uk/home/research/waveletcoherence/}
to get the XWT  of  the two time series in order to investigate
their phase relationship. Figure  4 shows the XWT  of the two time
series.  The two time series
are in phase in the area, namely about an 11 yr periodic belt, with
significant common power, because all arrows point right in the
area.

\begin{figure}
\hskip 10.mm
\includegraphics[width=70mm,height=80mm]{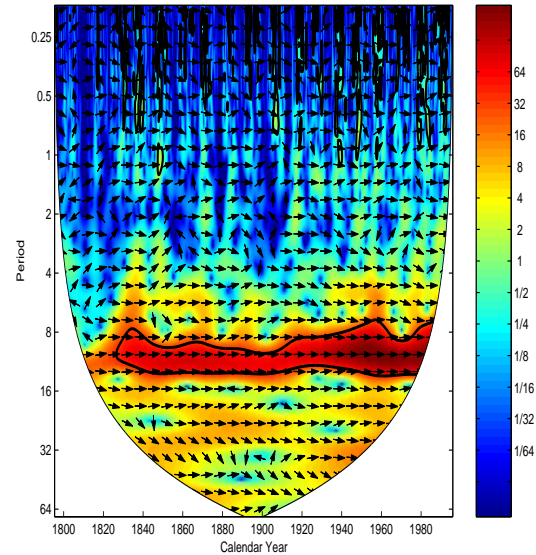}
\caption{The cross wavelet transform of the  monthly mean group
sunspot number and the monthly mean sunspot number. The thick black
contours  indicate the $95\%$ confidence
  level,  and the  region below the thin line indicates the
COI. The relative phase relationship is shown as arrows with
  in-phase pointing right,  anti-phase pointing left, and the former leading the latter  by $90^{\circ}$
  pointing straight down.
 }
\label{label4}
\end{figure}

The real and imaginary parts of a wavelet power spectrum are
considered separately, and the phase angle at any point of the
wavelet power spectrum can be obtained through its complex spectral
amplitude (Li et al. 2009). Figure  5 shows the phase angle
(argument) of the complex cross-wavelet  amplitudes varying with
frequency and time. It is the relative phase of the two time series.
The results shown in the figure clearly indicate that the relative
phases coherently fluctuate within a  rather large range of
frequencies, which correspond to the period scales of larger than 4
years. At smaller scales, there are no regular oscillatory patterns
in the corresponding frequency bands, and the high-frequency
components demonstrate a noisy behavior with strong phase mixing.

\begin{figure}
\hskip 10.mm
\includegraphics[width=70mm,height=80mm]{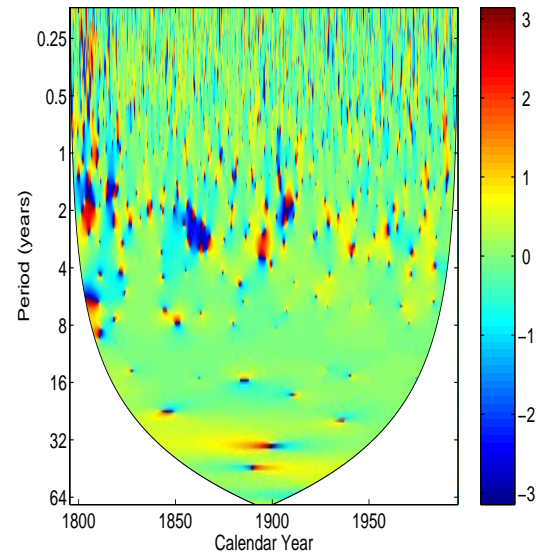}
\caption{The relative phase angles (arguments) of the complex
cross-wavelet amplitudes varying with frequency and time. The thin
solid line indicates the COI.
 }
\label{label5}
\end{figure}

Based on Figure 5, we calculate the averages of relative phase
angles over all scales for all time points of the considered
interval, which are displayed in Figure  6. Their corresponding
standard deviations are also calculated and shown in Figure  7. As
the two figures show, the relative phase of the two time series
fluctuates within larger amplitudes in cycles 5 and 6 than that in
all the following cycles does, and it fluctuates within larger
amplitudes usually around the maximum time of a cycle than around
the minimum time of the cycle. Asynchrony of the two time series is
thus more obviously seen in cycles 5 and 6 than in the following
cycles, and usually more obviously seen around the maximum time of a
cycle than during the rest of the cycle.

\begin{figure}
\hskip 10.mm
\includegraphics[width=70mm,height=80mm]{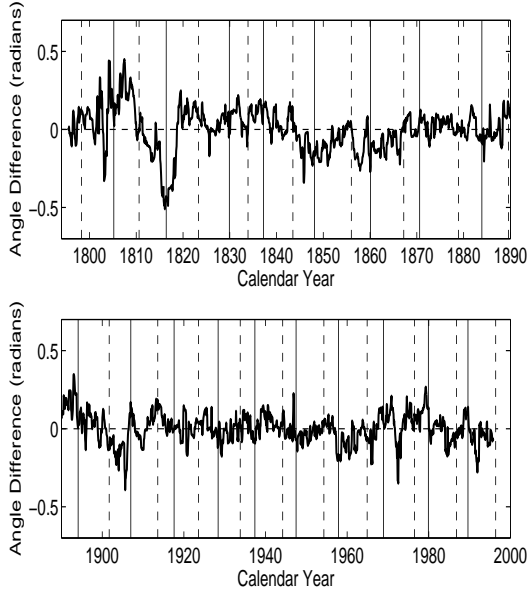}
\caption{ The average (the thick solid line) of relative phases over
all period scales shown in Figure 5 varying with
 time. The dashed (solid) vertical lines show the minimum (maximum) times of cycles.
 }
\label{label6}
\end{figure}

\begin{figure}
\hskip 10.mm
\includegraphics[width=70mm,height=80mm]{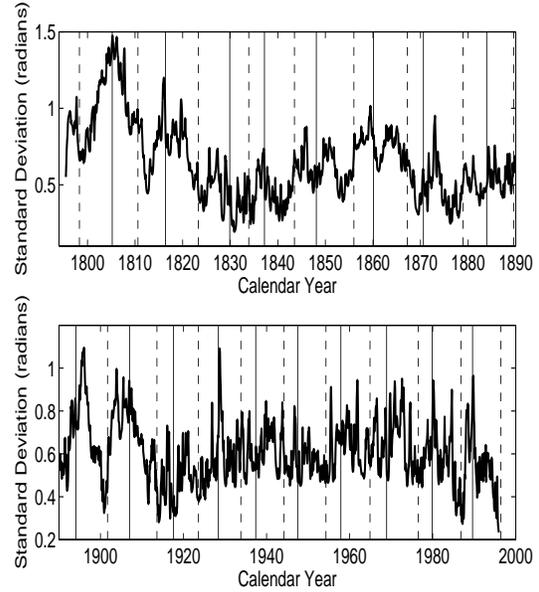}
\caption{The corresponding standard deviation (the thick line) of
the average of relative phases shown in Figure 6. The dashed (solid)
vertical lines show the minimum (maximum) times of cycles.
 }
\label{label7}
\end{figure}

For the two time series, the mean relative phases are shown in
Figure 8 in dependence on the considered frequencies. The
corresponding standard deviations have been calculated as well and
are given in the second panel of the figure. As the figure shows,
the relative phases  obviously coherently fluctuate  within time
scales of about 8 to 13 years (and even within 8 to 25 years),
 where their relative phases  are within $\pm0.2$  radian, and
their standard deviations are less than 0.5 radian. Although the
relative phases
 are always negative within time scales of about 8 to 13 years,
 seemingly indicating $R_{z}$
leading $R_{g}$, the corresponding standard deviations are so large
that we could not think $R_{z}$ leading $R_{g}$.  At time scales of
less than 8 years, the relative phase angles violently fluctuate and
their standard deviations are very large. The relative phase angles
increase after time scales of larger than 25 years, and their
standard deviations also increase.

\begin{figure}
\hskip 10.mm
\includegraphics[width=70mm,height=80mm]{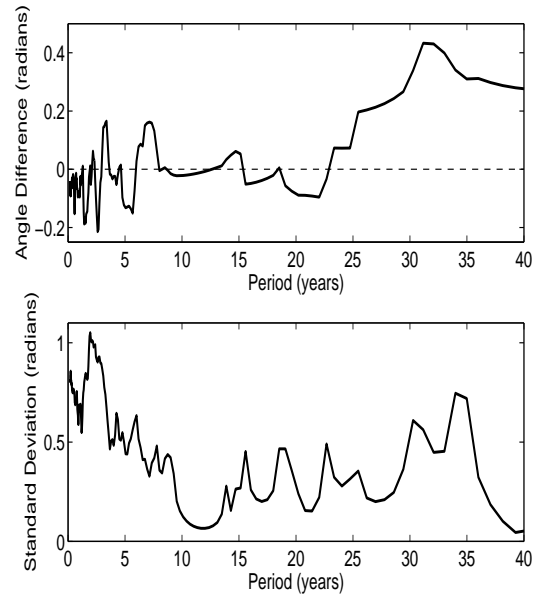}
\caption{Angle difference (the relative phase) of the XWT power
spectrum of of the monthly mean group sunspot number and the monthly
mean sunspot number as a function of periods (the top panel) and its
corresponding standard deviations (the bottom panel). Positive
values should be interpreted as the monthly mean group sunspot
number leading the monthly mean sunspot number.
 }
\label{label8}
\end{figure}

Taking very small or very large reference time scales into account
should lead to non-coherent behavior of the phase variables assigned
to $R_{g}$ and $R_{z}$. In the following, we therefore focus
attentions on periodicities
 in the coherent range
around the Schwabe cycle as the reference time scales. Because the
high-frequency components of $R_{g}$ and $R_{z}$ are filtered out in
such a case, the resulting phases are well-defined and may be used
to study the varying relationship between $R_{g}$ and $R_{z}$. Based
on Figure  5, we calculate the averages of relative phase angles
over scales of 8 to 13 years for all time points of the considered
interval, which are displayed in Figure  9. Their corresponding
standard deviations are also calculated and shown in the figure. As
the figure indicates, for all cycles except cycles 5 and 6, the mean
relative phase angles are very small (within $\pm0.35$ radian),
 fluctuating around the value of zero,
 and the standard deviations are small (less than 0.5 radian).
It is the low-frequency components of hemispheric flare activity in
period scales around the Schwabe cycle that  are responsible for the
strong phase synchronization. Based on Figure  5, we calculate the
averages of relative phase angles over scales of 8 to 25 years for
all time points of the considered interval, which are also displayed
in Figure  9. Their corresponding standard deviations are
calculated and shown in the figure. As the figure indicates, for all
cycles except cycles 5 and 6, the mean relative phase angles are
very small (even within $\pm0.2$ radian),
 fluctuating around the value of zero,
 and the standard deviations are small (less than 0.7 radian).
For cycles 5 and 6, there are no coherent frequencies.

\begin{figure}
\hskip 10.mm
\includegraphics[width=70mm,height=80mm]{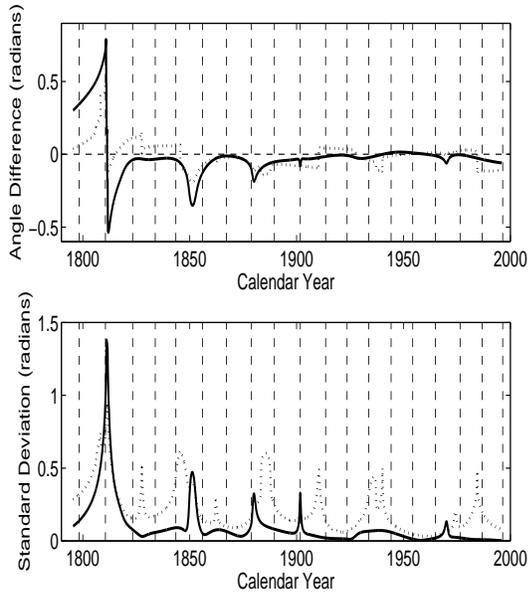}
\caption{The average of relative phases  over the period scales
respectively of 8 to 13 years (the solid line) and of 8 to 25  years
(the dotted line) varying with
 time (the top panel), and its corresponding standard deviation (the lower panel)
 with the XWT  method used. Solid (dotted) line corresponds to solid (dotted) line in the two panels.
 }
\label{label9}
\end{figure}

\section{Influence of their different definition on the relation between them}
As we know,
during the minimum time of a cycle a group of sunspots has only one
sunspot in the most cases or two sometimes, thus
$(10g+f)/g\approx11$. However, during the maximum time of a cycle a
group of sunspots has one to two sunspots sometimes, or several (or
even more) sunspots sometimes, thus $(10g+f)/g$ is larger than 11
and believed even usually larger than 12.08 (the normalization
constant in the definition of $R_{g}$). Therefore, there should certainly exist difference
between the two caused by their different definition. The median value of  $R_{g}$
is about 46.4, and $R_{g}$ is divided into two group: in one (G1) $R_{g}$ is less than,
and in the other (G2) $R_{g}$ is larger than
the median. A linear fit is done for each of the two groups, which is shown in Figure 10.
The fitting line is $R_{g}=1.5+0.897\times R_{z}$ with the correlation coefficient
of 0.965 for the group $G1$, and
$R_{g}=8.2+0.883\times R_{z}$ with the correlation coefficient of 0.962 for the group $G2$.
As the figure shows, the fitting line is very close to the straight line $R_{g}=R_{z}$ for $G1$.
However,  for $G2$
the fitting line is generally below the straight line $R_{g}=R_{z}$, and
the deviation between the two increases with the increase of $R_{g}$.
Conclusively, $R_{g}$ matches $R_{z}$ very well around the minimum time of a cycle,
but during the maximum time of a cycle it seems usually less than  $R_{z}$.
 $R_{g}$ and $R_{z}$ are doomed by  their different definition to present different relationship between them at different stage of a cycle.

\begin{figure}
\hskip 10.mm
\includegraphics[width=70mm,height=80mm]{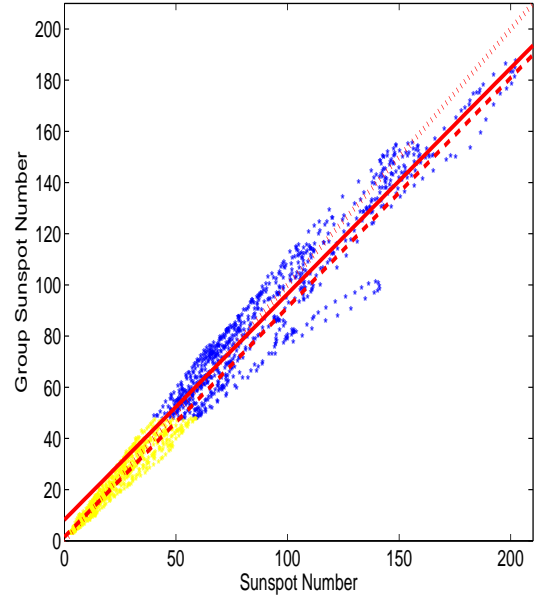}
\caption{Relationship  between $R_{g}$ and $R_{z}$.
$R_{g}$ is divided into two group: in one (G1) $R_{g}$ is less than its median (the yellow asterisks),
and in the other (G2) $R_{g}$ is larger than
the median (the blue asterisks). A linear fit is done for each of the two groups (the dashed line is for G1, and the solid for G2). The dotted line is the line of that $R_{g}=R_{z}$.
 }
\label{label10}
\end{figure}

The maximum value of a sunspot cycle is an important parameter to characterize the sunspot cycle.
Figure 11 displays the maximum values of the Wolf sunspot number  and
the group sunspot number in the modern cycles (cycles 10 to 22). The maximum value of the sunspot number in a cycle
is usually larger than that of the group sunspot number (cycles 10 to 12 and 18 to 22)
or very close to the corresponding one of the group sunspot number (cycles 13 to 17), confirming the aforementioned
result.  The average  of the sunspot number maxima over cycles 10 to 22 is 135.5, and
the average  of the group sunspot number maxima is 132.0, slightly less than the former. We also show in the figure the
difference  between the maximum  time of the sunspot number and
the group sunspot number in a cycle for the modern cycles. In 9 of the total 13 cycles, the difference is obvious,
indicating asynchrony of
$R_{g}$ and $R_{z}$.

\begin{figure}
\hskip 10.mm
\includegraphics[width=70mm,height=80mm]{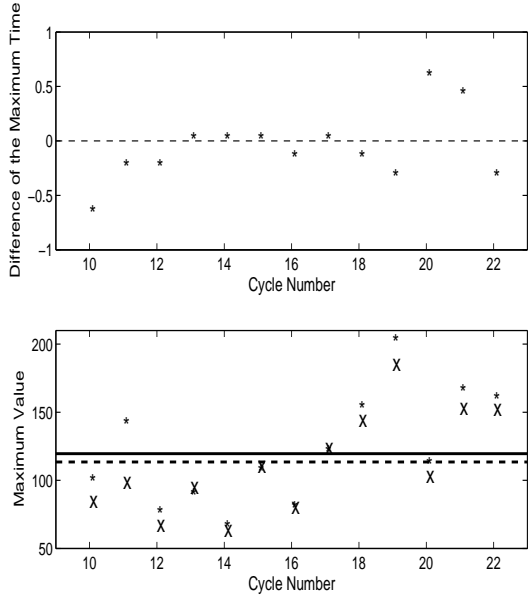}
\caption{The top panel: difference (asterisks) between the maximum  time of  the sunspot number and
the group sunspot number in a cycle. The dashed line shows no differences. The bottom panel: the maximum values of
the sunspot number (asterisks) and
the group sunspot number (crosses). The solid line shows the average (135.5) of the sunspot number maxima over cycles 10 to 22, and
the dashed, the average (132.0) of the group sunspot number maxima.}
\label{label11}
\end{figure}

\section{Conclusions and discussions}
 In the present study, the continuous wavelet
transform and the cross-wavelet transform have been proposed to
investigate the phase synchrony of the monthly mean Wolf and group
sunspot numbers, the longest direct instrumental records of solar
activity, and the data used here span from June 1795 to December
1995.

The continuous wavelet transforms of the two time series show that
the 11-year Schwabe cycle is the only one period of statistical
significance for the two time series, and
 it is 10.7 years.
However, in cycles 5 to 7 the Schwabe cycle does not significantly
exist for  $R_{g}$, and in cycles 5 and 6 for $R_{z}$.

The Schwabe period for $R_{g}$ and $R_{z}$ actually slightly differs
from each other before cycle 12, but from cycle 12 onwards  it is
almost the same for the two time series.
 If period (frequency) of two time series is different, the two are
asynchronous. The slight differences of the Schwabe cycle length for
the two time series should lead to a slight phase asynchrony between
them.

The cross-wavelet transform (XWT) of the two time series shows that
there is an area, locating at the Schwabe-cycle periodic belt, where
the two time series are approximately in phase. Through the XWT
analysis, the relative phase angles are found to coherently
fluctuate in a small angle range only within a range of frequencies
which corresponds to time scales around the Schwabe cycle, and even
within time scales of about 8 to 25 years. The high-frequency
components demonstrate a noisy behavior with strong phase mixing.
It is the low-frequency components of $R_{g}$ and $R_{z}$ in period
scales around the Schwabe cycle that are responsible for the strong
phase synchronization. Taking very small or very large reference
time scales into account should lead to non-coherent behavior of the
phase variables assigned to $R_{g}$ and $R_{z}$ at the respective
scales. Asynchrony of the two time series is found more obvious in
cycles 5 and 6 than in the following cycles, and more obvious
usually around the maximum time of a cycle than during the minimum
time of the cycle, indicating that $R_{g}$ should fit $R_{z}$ better in
both amplitudes and peak epoch during the minimum time time of a
solar cycle than during the maximum time of the cycle. As we know,
during the minimum time of a cycle a group of sunspots has only one
sunspot generally or two sometimes, thus
$(10g+f)/g\approx11$. However, during the maximum time of a cycle a
group of sunspots has one to two sunspots sometimes, or several (or
even more) sunspots sometimes, thus $(10g+f)/g$ is larger than 11
and believed even usually larger than 12.08 (the normalization
constant in the definition of $R_{g}$), and further it is scattered
within a wider range than during the minimum of the cycle. $R_{g}$ is
destined to
match $R_{z}$ better during the minimum time than during the
maximum time of a cycle, making the two more obviously asynchronous
during the maximum time than during the minimum time of a cycle,
and further, $R_{g}$ is usually less than $R_{z}$ around the maximum
time of a cycle.

Somewhat in agreement with the more obvious asynchrony of the two
during the maximum times than during the the corresponding minimum
times of solar cycles, {\it the ``Waldmeier Effect" - the
anti-correlation between cycle amplitude and the elapsed time
between minimum and maximum of a cycle - is much more apparent in
the Zurich numbers; the ``Amplitude-Period Effect" - the
anti-correlation between cycle amplitude and the length of the
previous cycle from minimum to minimum - is also much more apparent
in the Zurich numbers;  the ``Even-Odd Effect" - in which
odd-numbered cycles are larger than their even-numbered precursors -
is somewhat stronger in the Group numbers but with a tighter
relationship in the Zurich numbers; the `Secular Trend' - the
increase in cycle amplitudes since the Maunder Minimum - is much
stronger in Group numbers} (Hathaway, Wilson $\&$ Reichmann 2002).
Due to that $R_{g}$ matches $R_{z}$ worse during the maximum time
than during the minimum time of a cycle, they exhibit the above
different sunspot cycle characteristics and  more obvious asynchrony
during the maximum time of the cycle.

Sunspot number, group sunspot number and sunspot area are widely
and frequently utilized in astronomy and Earth science  to embody
long-term variations of solar activity. However, reviewed from the
phase relationship between each two of the three, they are slightly
asynchronous in phase, although they are coherent in the
low-frequency components corresponding to the  period scales around
the Schwabe cycle, this is inferred  the main reason why ``the
Zurich numbers follow the 10.7-cm radio flux and total sunspot area
measurements only slightly better than the Group numbers"
(Hathaway, Wilson $\&$ Reichmann 2002). Which one of the three can be served
as the best primary time series to define solar activity? it is an
open question, and careful attention should be paid.

\acknowledgements
We thank the anonymous referees for their careful reading of the
manuscript and constructive comments which improved the original version of the manuscript.
Data used here are all downloaded from web sites.
The authors would like to express their deep thanks to the staffs of
these web site.  The work is supported by the NSFC under Grants
10873032, 40636031, and 10921303, the National Key Research Science
Foundation
(2006CB806303), and the Chinese Academy of Sciences.

\newpage

\clearpage


\begin{thebibliography}{}
\bibitem{} Echer, E., Gonzalez, W.D., Guarnieri, F.L., Lago, A.Dal, Vieira, L.E.A.:
2005, Advances in Space Research 35, 855
\bibitem{} Eddy, J.A.: 1976, Science 192, 1189
\bibitem{} Faria, H.H., Echer, E., Rigozo, N.R.,   Viera, L.E.A.: 2004, Sol. Phys.
223, 305
\bibitem{} Grinsted, A., Moore, J.C.,  Jevrejeva, S.: 2004,  Nonlinear.
Proc. Geophys. 11, 561
\bibitem{} Hathaway, D.H.: 2010, Living Reviews in Solar Physics 7, 1
\bibitem{} Hathaway, D.H.,  Wilson, R.M.,
Reichmann, E.J.: 2002, Sol. Phys. 211, 357
\bibitem{} Hoyt, D.V.,   Schatten, K.H.: 1995a,
Sol. Phys. 160, 371
\bibitem{} Hoyt, D.V.,   Schatten, K.H.: 1995b,
Sol. Phys. 160, 379
\bibitem{} Hoyt, D.V.,  Schatten, K.H.: 1995c, Sol.
Phys. 160, 387
\bibitem{} Hoyt, D.V.,    Schatten, K.H.: 1995d, Sol. Phys. 160, 393
\bibitem{} Hoyt, D.V.,   Schatten, K.H.: 1997, The Role of Sun in
Climate Changes (New York: Oxford University Press)
\bibitem{} Hoyt, D.V.,  Schatten, K.H.: 1998a, Sol. Phys. 179, 189
\bibitem{} Hoyt, D.V.,  Schatten, K.H.: 1998b, Sol. Phys. 181, 491
\bibitem{} Kane, R.P.: 2002, Sol. Phys. 205, 383
\bibitem{} Li, K.J.,  Gao, P.X.,   Su, T.W.:   2005, Sol. Phys. 229, 181
\bibitem{} Li, K.J.,  Gao, P.X., Zhan, L.S., Shi, J.X.,   Zhu, W.W.:   2010, MNRAS, 401, 342
\bibitem{} Maraun, D., Kurths, J.: 2004, Nonlinear Proc. Geophys. 11, 505
\bibitem{} Ogurtsov, M.G.,
Nagovitsyn, Yu.A., Kocharov, G.E.,   Jungner, H.: 2002, Sol.
Phys. 211, 371
\bibitem{} Pikovsky, A., Rosenblum, M.,
Kurths, J.: 2001, Synchronization: A Universal Concept in Nonlinear
Science (Cambridge: Cambridge University Press)
\bibitem{} Rigozo, N.R., Echer, E., Vieira, L.E.A., Nordemann, D.J.R.: 2001, Sol. Phys. 203, 179
\bibitem{} Romano, M.C., Thiel, M., Kurths, J.,
Kiss, I.Z.,  Hudson, J.L.: 2005, Europhys. Lett. 71, 466.
\bibitem{} Torrence, C., Compo, G.P.: 1998,  Bull.  Amer. Meteor. Soc. 79, 61
\bibitem{} Wilson R.M.: 1998, Sol. Phys. 182, 217
\end{thebibliography}
\end{document}